\documentstyle[epsfig]{aipproc}
\def\Journal#1#2#3#4{{\em #1}{\bf #2}, #3 (#4)}

\def\PRD{Phys.~Rev. \bf{D}}

\begin{document}
\title{Existence of $\sigma$(600)/$\kappa$(900)-Particle
and New Chiral Scalar Nonet\ ``Chiralons''}

\author{
Muneyuki Ishida$^*$ and Shin Ishida$^{\dagger}$}
\address{
$^*$
Department of Physics, University of Tokyo, Tokyo 113 and
$^{\dagger}$
Atomic Energy Research Institute, College of Science and Technology,
Nihon University, Tokyo 101, JAPAN}

\maketitle

\begin{abstract}
The $\sigma$(600) and $\kappa$(900),
observed in the phase shift analyses\cite{rf:had97,rf:had97a}, 
satisfy rather well the mass and width relation predicted by 
the SU(3)L$\sigma$M and the SU(3)L$\sigma$M with the 
vector and axial-vector meson nonets, and 
 deserve to be the members of scalar $\sigma$-nonet,
together with the observed resonances $a_0(980)$ and $f_0(980)$, 
 as a chiral partner of pseudoscalar $\pi$-nonet.
In the phase shift analyses an introduction of 
repulsive background phase $\delta_{BG}$ is essential, 
whose origin has a close connection to the 
$\lambda\phi^4$ interaction in L$\sigma$M.
It is argued that the members of this $\sigma$-nonet,
``Chiralons," have different properties and should
be discriminated from the conventional 
$^3P_0$-$q\bar{q}$-scalar nonet.
\end{abstract}


 ({\em Properties of 
$\sigma /\kappa$ and Linear $\sigma$ Model})\ \ \ \ \ 
The $\sigma (600)$ and $\kappa (900)$
are observed by the reanalyses\cite{rf:had97,rf:had97a} 
of $\pi\pi$ and $K\pi$ phase shifts,
respectively. 
Here their properties in relation to the 
 SU(3) chiral symmetry is examined. 
In the low energy region, where the 
structure of composite meson can be neglected, 
 the 
 linear $\sigma$ model(L$\sigma$M) may be valid.
In terms of matrix 
notation $B\equiv\sigma +i\phi$(
$\sigma\equiv\lambda^i\sigma^i/\sqrt{2}$ and 
$\phi\equiv\lambda^i\phi^i/\sqrt{2}$
denoting scalar and pseudoscalar meson nonet, respectively),
the Lagrangian 
of $SU(3)$L$\sigma$M\cite{rf:gg,rf:CH} is
\begin{eqnarray}
{\cal L} &=& 
\frac{1}{2}\langle \partial_\mu B\partial^\mu B
^\dagger 
\rangle 
-\frac{\mu^2}{2}\langle BB
^\dagger 
\rangle
-\frac{\lambda_1}{4}\langle BB^\dagger \rangle^2
-\frac{\lambda_2}{2}\langle (BB^\dagger )^2\rangle
\nonumber\\  
 &+& \kappa_d({\rm det}\ B+{\rm det}\ B^\dagger )
+\langle f \sigma\rangle ,
\label{eq:lsm}
\end{eqnarray}
where $\langle\ \ \rangle$ represents the trace,
and the form of  
 $f(=diag\{ f_n,f_n,f_s\} )$ guarantees the PCAC.
The six model parameters
contained in Eq.(\ref{eq:lsm}) 
are determined by
the masses of $\pi ,\eta , \eta '$, $\sigma$ and  $\kappa$,
and the decay constant
$f_\pi$, and then
we can predict the properties of scalar mesons, given 
in Table \ref{tab:tree}:
\footnote{The predicted properties of scalar mesons 
are very sensitive to the value of 
$f_K/f_\pi$\cite{rf:CH}. 
I take the region of 
this ratio $1.329<f_K/f_\pi<1.432$,
where the $m_\kappa^{\rm theor}$ 
reproduces the experimental value 
within its uncertainty,
being somewhat larger than the experimental one,
$f_K/f_\pi= 113/93=1.22$.}
\begin{table}
\begin{center}
\caption{The masses and widths of scalar meson nonet(``chiralons") 
predicted by SU(3)L$\sigma$M, compared with experiments
(the unit is in MeV). 
The values with underlines are used as inputs.
The $\Gamma_{\sigma '}^{\rm theor}$ is the $\pi\pi$-width.
}
\begin{tabular}{|c|c|c|c|c||c|c|c|c|c|}
\hline  
   & $m^{\rm theor}$ & $m^{\rm exp}$
   & $\Gamma^{\rm theor}$ & $\Gamma^{\rm exp}$ &
   & $m^{\rm theor}$ & $m^{\rm exp}$
   & $\Gamma^{\rm theor}$ & $\Gamma^{\rm exp}$\\
\hline
  $\sigma$ & \scriptsize\underline{535$\sim$650} & 
             \scriptsize\underline{535$\sim$650}
           & \scriptsize{400$\sim$800} & \scriptsize{385$\pm$70} &
  $\delta$ & \scriptsize{900$\sim$930} & \scriptsize{984$\pm$1}  
           & \scriptsize{110$\sim$170} & \scriptsize{50$\sim$100} \\
\hline
  $\kappa$ & \underline{\scriptsize{905}$\stackrel{+65}{\scriptstyle -30}$}
           & \underline{\scriptsize{905}$\stackrel{+65}{\scriptstyle -30}$}  
   & \scriptsize{300$\sim$600} 
   & \scriptsize{$545\stackrel{+235}{\scriptstyle -110}$} &
  $\sigma '$ & \scriptsize{1030$\sim$1200}  & \scriptsize{980$\pm$10}
             & \scriptsize{0$\sim$300}   & \scriptsize{40$\sim$100} \\
\hline
\end{tabular}
\label{tab:tree}
\end{center}
\end{table}
The predicted widths of $\sigma$ and $\kappa$
are consistent to the experimental values. 
Thus $\sigma (600)$ and $\kappa (900)$ 
have plausible properties as the members of
the scalar nonet in L$\sigma$M.
The predicted properties of the other members,
$\delta (I=1)$ and $\sigma '(s\bar{s})$, are
close to those of
$a_0(980)$ and
$f_0(980)$. 
Thus, it may be plausible to regard 
{\bf the}  
${  \sigma}${\bf (600)}, ${  \kappa}${\bf (900)}, 
{\bf a}${}_0${\bf (980)},{\bf and} {\bf f}${}_0${\bf (980)}
{\bf as being the members of the scalar nonet
forming with the members of} ${  \pi}${\bf -nonet the  linear 
representation of the SU(3) chiral symmetry}.

 ({\em repulsive core and $\lambda\phi^4$ term})\ \ \ \ \ 
In our phase shift analyses the repulsive 
background phase shift of
hard core type 
$\delta_{\rm BG}$ introduced phenomenologically 
plays an essential role. The 
origin of this $\delta_{\rm BG}$ seems to  have a close connection to 
 the $\lambda\phi^4$-interaction in L$\sigma$M\cite{rf:MY}:
\begin{figure}
 \epsfysize=4.5 cm
 \centerline{\epsffile{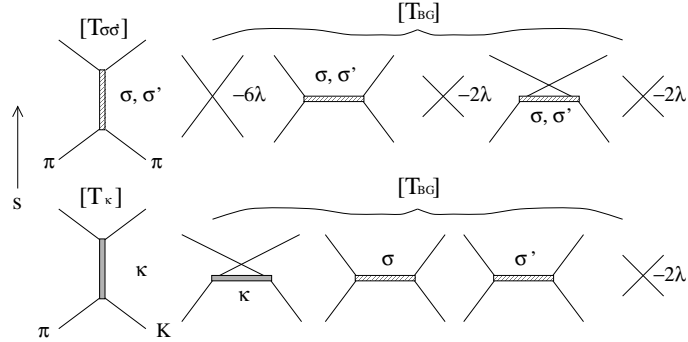}}
\caption{$\pi\pi$ and $K\pi$ scattering diagram
in L$\sigma$M}
\label{fig:diagram}
\end{figure}
It represents a contact zero-range interaction
and is strongly repulsive,
and has plausible properties
as the origin of repulsive core.
In the following we shall make a tentative
theoretical estimate of the core radii
in the framework of L$\sigma$M:
The diagrams for 
the $\pi\pi$ and $K\pi$ scattering amplitudes 
in the tree level,
are given in Fig. \ref{fig:diagram}.
The scattering ${\cal T}$ matrix consists of a resonance part
${\cal T}_R$ and of a background part ${\cal T}_{BG}$.
The ${\cal T}_{BG}$ corresponds to the contact
$\lambda\phi^4$ interaction and the exchange of
relevant resonances.
This ${\cal T}_{BG}$ has a weak $s$-dependence in comparison
with that of ${\cal T}_{R}$.
We regard them as ${\cal K}$-matrix (${\cal K}_R\equiv {\cal T}_R$, 
${\cal K}_{BG}\equiv {\cal T}_{BG}$,)
and 
unitarize them separately by
$S=\prod_{A}S_A$\cite{rf:had97a};\ \ \ $S_A=
(1+i\rho_1{\cal K}_A)/(1-i\rho_1{\cal K}_A)\ \ (A=R,BG)$.
We estimate the theoretical core radius at 
the resonance energy by the formula,
$r_c^{\rm theor}\equiv\frac{1}{p_1}
{\rm atan}(-\rho_1{\cal T}_{BG})|_{\sqrt{s}=m_R}$, and the results 
 are given in Table \ref{tab:rc}.
They seem almost consistent to the experimental values
obtained in the phase shift analysis\cite{rf:had97a}.

\begin{table}[t]
\caption{
Repulsive core radii $r_c$(fm) in $\pi\pi$ and $K\pi$ scattering
}
\begin{center}
\begin{tabular}{|l|c|c|c|c|}
\hline
   & $(I=0)_{\pi\pi}$ & $(I=1/2)_{K\pi}$ 
   & $(I=2)_{\pi\pi}$ & $(I=3/2)_{K\pi}$\\
\hline
 experiment & 0.60$\pm$0.07 & 0.70$\pm$0.08 & 0.17 & 0.16\\
\hline
 theory(L$\sigma$M) & 0.6$\sim$0.8
 & 0.3$\sim$0.4
 & 0.13$\sim$0.17
 & 0.11$\sim$0.15 \\
\hline
\end{tabular}
\label{tab:rc}
\end{center}
\end{table}

 ({\em Why $\sigma$ and $\kappa$ 
have been overlooked})\ \ \ \ 
According to the above analysis  
the $\sigma (\sigma ')$ consists almost of 
$n\bar n$-($s\bar s$-)component, and I take the SU(2)L$\sigma$M
 for simplicity. 
The $\pi\pi$-scattering $A(s,t,u)$-amplitude is given by \ 
$
A(s,t,u) = (-2g_{\sigma\pi\pi})^2/(m_\sigma^2-s)-2\lambda .
$
Because of the relation,
$
g_{\sigma\pi\pi} =
f_\pi\lambda =(m_\sigma^2-m_\pi^2)/(2f_\pi ), 
$
the amplitude due to 
virtual $\sigma$ production( 1st term)
is exactly 
cancelled by that due to 
repulsive $\lambda\phi^4$ interaction( 2nd term) 
in $O(p^0)$ level, 
and 
the $A(s,t,u)$ is rewritten into the following form:
\begin{eqnarray}
A(s,t,u) &=& \frac{1}{f_\pi^2}
\frac{(m_\sigma^2-m_\pi^2)^2}{m_\sigma^2-s}
-\frac{m_\sigma^2-m_\pi^2}{f_\pi^2}
=\frac{s-m_\pi^2}{f_\pi^2}+\frac{1}{f_\pi^2}
\frac{(m_\pi^2-s)^2}{m_\sigma^2-s},
\label{eq:Acancel}
\end{eqnarray}
where
 in the last side 
the $O(p^2)$ Tomozawa-Weinberg amplitude
and the $O(p^4)$ (and higher order) correction 
term are left.
As a result the derivative coupling property
of $\pi$-meson as a Nambu-Goldstone boson is
preserved.
In this sense the $\lambda\phi^4$-interaction can be
called a ``compensating" interaction for
$\sigma$-effect.

The essential point in our phase shift analyses
is a strong cancellation
between the positive $\delta_\sigma$ ($\delta_\kappa$) 
and the negative $\delta_{\rm BG}$,
as shown in Table 1 in the contribution to this 
conference\cite{rf:had97}, which 
 corresponds to the 
Eq.(\ref{eq:Acancel}) in L$\sigma$M.

In the conventional 
Breit-Wigner formula a 
non-derivative $\sigma\pi\pi$-coupling,
accordingly a non-derivative
effective $\pi\pi$-interaction, is supposed without
consideration on the ``compensating" repulsive interaction.
This seems\cite{rf:MY} to be a reason why 
$\sigma$ (and $\kappa$) has been overlooked  
in most of the $\pi\pi$\cite{rf:MP} (and $K\pi$) phase shift
analyses.

 ({\em Related problems})\ \ \ \ \ 
An investigation of the properties 
 of the scalar $\sigma$-nonet with a wide scope is done 
 by using  the ``super'' 
Lagrangian\cite{rf:gg}, where  
 the vector($\rho$-)
and the  axial-vector($a_1$-) meson nonets
are included consistently to  
 the vector meson dominance
. 
The eight new parameters
in addition to the six parameters
in Eq.(\ref{eq:lsm})
are appeared in this scheme. 
 In Table  \ref{tab:super},
 the theoretical values of 21 quantities( including $f_\pi$ and $f_K$)
are given in comparison with the experimental values. 
The obtained values of scalar meson properties are 
 quite close to those in the original L$\sigma$M,
Eq.(\ref{eq:lsm}).
Thus now it becomes 
more plausible than before that   
the $\sigma$(600), $\kappa$(900), $\delta$(980)(=$a_0$(980)) 
and $\sigma '$(980)(=$f_0$(980)) form  
 the chiral scalar-nonet, which was first 
 suggested in ref. \cite{rf:sca}.

\begin{table}
\caption{
Properties of scalar, pseudoscalar and vector mesons
 by the L$\sigma$M with $\rho$- and $a_1$-nonets.
(a) Masses and widths(MeV) of scalar mesons.
(b) Masses(MeV) of pseudoscalar and vector mesons
and  widths(MeV) of vector mesons.
The $f_\pi (f_K)$ is obtained with 93.2(114.7)MeV,
 consistent to the experimental value
.
}
\begin{center}
\begin{tabular}{|l|c|c|c|c||c|c|c|c|}
\hline
(a) & $m_\sigma$ & $m_\kappa$ & $m_\delta$ & $m_{\sigma '}$  
 & $\Gamma_\sigma$ & $\Gamma_\kappa$
 & $\Gamma_\delta$ & $\Gamma_{\sigma '}$ \\ 
\hline
theor & 600 & 989 & 955 & 1063 & 454 & 505 & 174 & 131 \\
exp & 535$\sim$675 & 905$\stackrel{+65}{\scriptstyle -30}$ 
 & 984$\pm$1 & 980$\pm$10 & 385$\pm$70 
 & 545$\stackrel{+235}{\scriptstyle -110}$ & 50$\sim$100 & 40$\sim$100\\
\hline
\end{tabular}

\begin{tabular}{|l|c|c|c|c||c|c|c|c||c|c|c|}
\hline
(b) & ~$m_\pi$~ & ~$m_K$~ & ~$m_\eta$~ & ~$m_{\eta '}$~  &
 ~$m_\rho$~ & ~$m_{K^*}$~ & ~$m_\omega$~ & ~$m_\phi$~
 & ~$\Gamma_\rho$~ & ~$\Gamma_{K^*}$~ & ~$\Gamma_\phi$~ \\
\hline
theor & 139 & 502 & 559 & 1020
 & 783 & 887 & 778 & 1021
 & 150 & 57 & 3 \\
exp & 138 & 495 & 547 & 958 
 & 769 & 894 & 782 & 1019
 & 151 & 50 & 4  \\
\hline
\end{tabular}
\label{tab:super}
\end{center}
\end{table}


It should be noted that the $\sigma$ is now directly observed in 
the production processes as shown in the contribution to this 
conference\cite{rf:had97d}.
Thus, the conventional argument\cite{rf:GL} against 
 the validity of L$\sigma$M and the existence of
$\sigma$-meson in the framework of 
chiral perturbation theory(ChPT),
which   
%
is  based only on the indirect experiments  concerning
a much lower energy region than  $m_\sigma$, is not appropriate. 
The parameters in L$\sigma$M are determined  
through the physics in resonance energy region,
and the related low energy quantities are predicted  
with no free fitting parameters, 
quite in contarst with the situation in ChPT.
For example,  the  $\pi\pi$-scattering lengths ${a_J}^I$
 obtained by
 the L$\sigma$M with $\rho$- and $a_1$-nonets 
are \ 
$({a_0}^0,{a_0}^2;{a_1}^1;{a_2}^0,{a_2}^2;{a_3}^1)
=(0.16,-0.010;0.022;13\times 10^{-4},2.3\times 10^{-4};
4\times 10^{-5})$,
which 
are not so much different from the present 
experimental values
. 
In this connection we should like to refer to 
the work\cite{rf:shaba} where 
the width and form factor of $K_{l4}$-decay,
as well as the $\pi\pi$-phase shift $\delta_0^0$
in low energy region,
are also well described by L$\sigma$M
similarly as in ChPT.\footnote{
It is discussed\cite{rf:meissner} that  
the effect of $\sigma$-meson can be taken into account through
the $O(p^4)$ and $O(p^6)$ Lagrangian in ChPT.
Such a non-linear
 approach 
has no predictive power
about the properties of the 
 scalar $\sigma$-meson nonet
as a chiral partner of the $\pi$-nonet, such 
as shown in Tables \ref{tab:tree} and \ref{tab:super}. }

Finally it is also to be noted that 
 the members of the $\sigma$-nonet 
(the $a_1$-nonet), 
``Chiralons," have different properties and should
be discriminated from the conventional 
$^3P_0$-($^3P_1$-)$q\bar{q}$-scalar nonet. (See, the contribution\cite{rf:had97}.)

\vspace*{-0.1cm}

\end{document}